\documentclass[tighten,twocolumn,apjl]{likeapj}
\usepackage{apjfonts}
\usepackage{amsmath}

\shortauthors{Fu}
\received{December 18, 2023}
\revised{January 21, 2024}
\accepted{February 10, 2024}
\published{}

\newcommand{\kms}{{km s$^{-1}$}}
\newcommand{\HI}{H\,{\sc i}}

\begin{document}

\title{Restoration of the Tully-Fisher Relation by Statistical Rectification}
\author{Hai~Fu}
\affiliation{Department of Physics \& Astronomy, University of Iowa, Iowa City, IA 52242}

\begin{abstract}

I employ the Lucy rectification algorithm to recover the inclination-corrected distribution of local disk galaxies in the plane of absolute magnitude ($M_i$) and \HI\ velocity width ($W_{20}$). By considering the inclination angle as a random variable with a known probability distribution, the novel approach eliminates one major source of uncertainty in studies of the Tully-Fisher relation: inclination angle estimation from axial ratio. Leveraging the statistical strength derived from the entire sample of 28,264 \HI-selected disk galaxies at $z < 0.06$ from the Arecibo Legacy Fast ALFA (ALFALFA) survey, I show that the restored distribution follows a sharp correlation that is approximately a power law between $-16 > M_i > -22$: $M_i = M_0 - 2.5\beta \ [\log(W_{\rm 20}/250 {\rm km/s})]$, with $M_0 = -19.77\pm0.04$ and $\beta = 4.39\pm0.06$. At the brighter end ($M_i < -22$), the slope of the correlation decreases to $\beta \approx 3.3$, confirming previous results. Because the method accounts for measurement errors, the intrinsic dispersion of the correlation is directly measured: $\sigma(\log W_{20}) \approx 0.06$\,dex between $-17 > M_i > -23$, while $\sigma(M_i)$ decreases from $\sim$0.8 in slow rotators to $\sim$0.4 in fast rotators. The statistical rectification method holds significant potential, especially in the studies of intermediate-to-high-redshift samples, where limited spatial resolution hinders precise measurements of inclination angles.

\end{abstract}


\section{Introduction} \label{sec:intro}

In observational astronomy, a prevalent challenge involves recovering intrinsic properties from observed ones. This restoration is essential due to the potential alteration of intrinsic properties by factors like viewing angles, dust extinction, atmospheric seeing, instrumental point spread function (PSF), as well as statistical and instrumental noise. In general, the transformation from the intrinsic probability distribution to the observed one follows the {\it law of total probability}, which for two-dimensional problems is:
\begin{equation} \label{eq:general}
\phi(x,y) = \iint \psi(\xi,\eta) P(x,y|\xi,\eta) d\xi d\eta, 
\end{equation}
where $(\xi, \eta)$ are the intrinsic properties, $(x,y)$ the observed properties, $\psi(\xi,\eta)$ the probability density function (PDF; or distribution in short) of the intrinsic properties, $\phi(x,y)$ the PDF of the observed properties, and $P(x,y|\xi,\eta)$ the conditional PDF. The definition of the conditional PDF is that $P(x,y|\xi,\eta) dx dy$ is the probability that $x'$ and $y'$ respectively fall in the interval $(x,x+dx)$ and $(y,y+dy)$ when $\xi' = \xi$ and $\eta' = \eta$. The objective is to reverse this equation to restore $\psi(\xi,\eta)$ from $\phi(x,y)$ given the knowledge of $P(x,y|\xi,\eta)$.

{\it Convolution} is a special case of Eq.\,\ref{eq:general} when the conditional PDF can be expressed as a function of the differences between intrinsic and observed properties:
\begin{equation} \label{eq:special}
\phi(x,y) = \iint \psi(\xi,\eta) K(x-\xi,y-\eta) d\xi d\eta,
\end{equation}
where $K(x-\xi,y-\eta)$ is called the convolution kernel, and its shape is invariant across the plane of $\xi$ and $\eta$. Similar to the general reversal problem, the goal of {\it deconvolution} is to recover $\psi(\xi,\eta)$ from $\phi(x,y)$ given the knowledge of $K(x-\xi,y-\eta)$. One particularly important problem is ``image restoration'', that is to recover the intrinsic surface brightness map by removing or reducing of the effects of atmospheric seeing and/or instrumental PSF due to telescope/interferometer geometry. Popular {\it deconvolution} algorithms include {\sc Richardson-Lucy} \citep{Richardson72,Lucy74}, {\sc Clean} \citep{Hogbom74,Cornwell09}, and {\sc Wiener-Hunt} \citep{Orieux10}. 

Because in general the conditional PDF $P(x,y|\xi,\eta)$ varies in the plane of $\xi$ and $\eta$, the aforementioned deconvolution algorithms cannot handle the reversal of Eq.\,\ref{eq:general}. But there is one exception. The iterative rectification algorithm of \citet{Lucy74} is in fact {\it designed} to reverse Eq.\,\ref{eq:general}, and that is the main difference between the \citet{Lucy74} rectification algorithm and the \citet{Richardson72} deconvolution algorithm. 

In this work, I employ the Lucy rectification algorithm to restore the Tully-Fisher relation \citep[TFR;][]{Tully77}, which is an important scaling relation between rotation velocity and luminosity of disk galaxies. In previous studies of the relation \citep[e.g.,][]{Tully12,Zaritsky14a,Tiley16,Desmond17,Ubler17,Topal18,Kourkchi22,Ball23}, the observed luminosities and velocity widths were corrected using inclination angles estimated from the observed axial ratios ($b/a$). The rectification method implemented here replaces the individual inclination correction with robust statistical rectification. By eliminating the reliance on axial ratio measurements and their conversion to inclination angles, this method removes a major source of error in determining the TFR.

The {\it Letter} is organized as follows. First, I introduce the rectification algorithm in \S\ref{sec:algorithm}. Next in \S\ref{sec:tfr}, I describe the survey data set (\S\ref{sec:phi}), construct the joint PDF (\S\ref{sec:PDF}), restore the intrinsic distribution (\S\ref{sec:psi}), and compare the resulting TFR with that from the $b/a$-based individual correction method (\S\ref{sec:comparison}). Finally, I summarize the work and comment on future applications of the method in \S\ref{sec:summary}.

\section{Algorithm} \label{sec:algorithm}

For simplicity, I first derive the equations of the algorithm in one dimension (1D), then provide the equivalent iterative equations for two-dimensional (2D) problems. 

The iterative rectification algorithm of \citet{Lucy74} is constructed using Bayes' theorem. Given the law of total probability, $\phi(x) = \int \psi (\xi) P(x|\xi) d\xi$, one can define its inverse integral, $\psi(\xi) = \int \phi(x) Q(\xi|x) dx$, with the inverse conditional PDF $Q(\xi|x)$. In other words, given that $P(x|\xi)dx$ is the probability that $x'$ falls in the interval $(x,x+dx)$ under the condition that $\xi' = \xi$, $Q(\xi|x)d\xi$ is the probability that $\xi'$ falls in the interval $(\xi,\xi+d\xi)$ under the condition that $x' = x$. With Bayes' theorem, $\phi(x) Q(\xi|x) = \psi(\xi)P(x|\xi)$, one can replace $Q(\xi|x)$ in the inverse integral and obtain the following equation:
\begin{align}
\psi(\xi) &= \int \phi(x) Q(\xi|x) dx \nonumber \\ 
	&= \int \phi(x) \Big( \frac{\psi(\xi)P(x|\xi)}{\phi(x)} \Big) dx \nonumber \\  
	&= \psi(\xi) \int \frac{\phi(x)}{\phi(x)} P(x|\xi) dx \label{eq:identity}
\end{align}
At first glance, the above may seem to be trivial as the terms cancel out and the integral of $P(x|\xi)$ must be unity by definition. But it inspired a highly efficient algorithm that allows the iterative solution of the intrinsic distribution function $\psi(\xi)$ from the observed distribution function $\phi(x)$, when the conditional PDF $P(x|\xi)$ is known. At the $r$-th iteration, the \citet{Lucy74} algorithm is simply described by two iterative equations:
\begin{align}
\phi^r(x) &= \int \psi^r(\xi) P(x|\xi)d\xi \label{eq:1d_iter_a} \\
\psi^{r+1}(\xi) &= \psi^r(\xi)\int \frac{\tilde{\phi}(x)}{\phi^r(x)} P(x|\xi) dx \label{eq:1d_iter_b}
\end{align}
Evidently, Eq.\,\ref{eq:1d_iter_a} is the law of total probability, and Eq.\,\ref{eq:1d_iter_b} is the iterative version of the Bayes identity in Eq.\,\ref{eq:identity}. For two-dimensional problems incorporating measurement errors ($\sigma_x, \sigma_y$), the iterative equations become:
\begin{align}
\phi^r(\tilde{x},\tilde{y}) &= \iint \psi^r(\xi,\eta) P(\tilde{x},\tilde{y}|\xi,\eta,\sigma_x,\sigma_y) d\xi d\eta \label{eq:phi_r} \\
\psi^{r+1}(\xi,\eta) &= \psi^r(\xi,\eta) \iint \frac{\tilde{\phi}(\tilde{x},\tilde{y})}{\phi^r(\tilde{x},\tilde{y})} P(\tilde{x},\tilde{y}|\xi,\eta,\sigma_x,\sigma_y) d\tilde{x} d\tilde{y} \label{eq:psi_r}
\end{align}

In Eqs.\,\ref{eq:1d_iter_b} \& \ref{eq:psi_r}, the key input is $\tilde{\phi}$, which is the observed distribution function of the observables. The tilde hat is used to distinguish it from the true distribution function, which is denoted simply as $\phi$. Similarly, I have used $(\tilde{x},\tilde{y})$ to denote the measured values (i.e., with errors) and $(x,y)$ to denote the true values (i.e., without errors).

Given the iterative equations, the procedure to carry out the iterative algorithm is to:
\begin{enumerate}
\item quantify the observed distribution function $\tilde{\phi}(\tilde{x},\tilde{y})$ from the data,
\item define the conditional PDF $P(\tilde{x},\tilde{y}|\xi,\eta,\sigma_x,\sigma_y)$ for the particular problem, 
\item prescribe an initial guess of the intrinsic distribution function $\psi^0(\xi,\eta)$, and 
\item solve iteratively for the intrinsic distribution function $\psi(\xi,\eta)$ with the previous two equations.
\end{enumerate}

\section{Application to the TFR} \label{sec:tfr}

In this section, I apply the iterative rectification method to restore the TFR from the observed distribution of Arecibo \HI-selected galaxies. A Python notebook of the full analysis is made publicly available\footnote{\url{https://github.com/fuhaiastro/TFR_Lucy}}. I define $\tilde{x}$ as the observed projected velocity width ($W_{20}$ in \S\ref{sec:phi}), $\xi$ the edge-on velocity width, $\tilde{y}$ the observed projected $i$-band absolute magnitude ($M_i$ in \S\ref{sec:phi}), and $\eta$ the face-on $i$-band absolute magnitude. The goal is to recover the distribution of the galaxy sample in $\xi$ and $\eta$, $\psi(\xi,\eta)$, from the observed distribution in $\tilde{x}$ and $\tilde{y}$, $\tilde{\phi}(\tilde{x},\tilde{y})$. As in the previous section, $x$ and $y$ are reserved for true projected velocity width and true projected $i$-band absolute magnitude in the absence of measurement errors, and they will be integrated out when evaluating the joint PDF of $\tilde{x}$ and $\tilde{y}$. 

\subsection{Data} \label{sec:phi}

\begin{figure}[!tb]
\epsscale{1.19}
\plotone{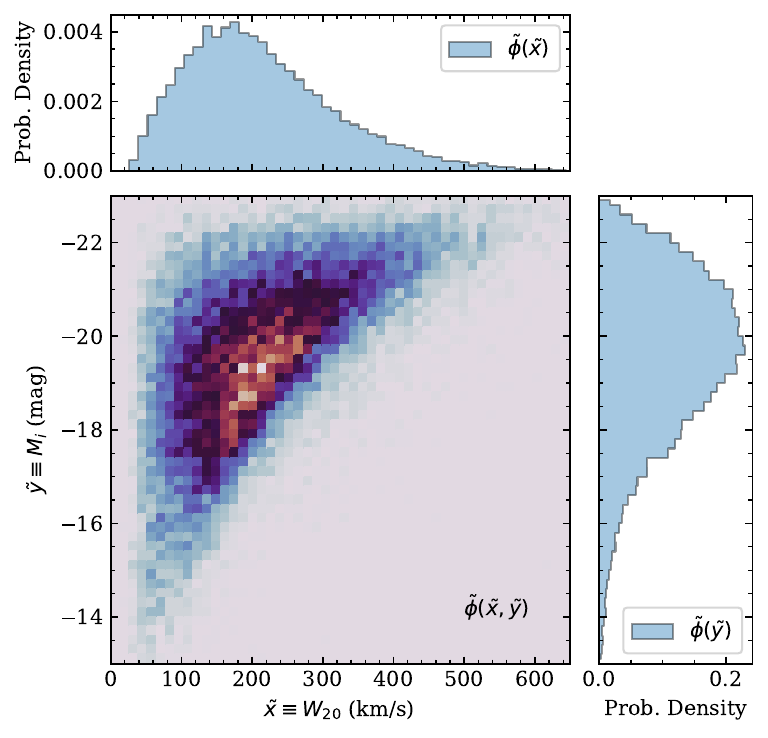}
\caption{The distribution of ALFALFA-SDSS galaxies in the plane of $i$-band absolute magnitude vs. \HI\ line width. This is the input $\tilde{\phi}(\tilde{x},\tilde{y})$ function, which will be rectified by the Lucy algorithm to {\it statistically} remove the effects from random disk orientations and random measurement errors. Here and in most of the subsequent figures, the main panel shows the 2D distribution, and the side panels show the marginalized distributions over each axis. 
\label{fig:phi}} 
\epsscale{1.0}
\end{figure} 

The \HI\ measurements are taken from the 100\% complete ALFALFA catalog \citep[the $\alpha$.100 sample;][]{Haynes18} and the absolute magnitudes are from the cross-matched ALFALFA-SDSS galaxy catalog \citep{Durbala20}. In both catalogs, distances to the galaxies are inferred from Hubble's law with $H_0 = 70~{\rm km~s}^{-1}~{\rm Mpc}^{-1}$ and a local peculiar velocity model \citep[for details, see \S3 of ][]{Haynes18}. I merge the two catalogs based on the Arecibo General Catalog (AGC) ID, resulting in a total of 31,500 entries with 53 columns. A total of 28,264 sources (90\% of the $\alpha$.100 sample) in the merged catalog have valid velocity widths and absolute magnitudes, and that forms the galaxy sample for this study, because no further down-selection is necessary. The sample is at low redshift ($z < 0.06$) and covers a wide range in $i$-band absolute magnitude ($-13 < M_i < -23$).

For the \HI\ velocity width, I start with the reported velocity width at 20\% level of each of the two peaks in the line profile (Column \texttt{W20}), because it is expected to capture more of the flat parts of a rotation curve than the 50\% level velocity width (\texttt{W50}). All reported velocity widths are corrected for instrumental broadening following the simulations of \citet{Springob05}. \texttt{W20} are given in observed frame instead of in rest frame, and the rest-frame \HI\ velocity widths ($W_{\rm 20}$) are obtained by dividing \texttt{W20} by $(1+cz_\odot/c)$, where $cz_\odot$ is the Heliocentric velocity of the \HI\ profile (column \texttt{Vhelio} in the catalog). 

For the $i$-band absolute magnitude, I start with the extinction corrected $i$-band absolute magnitude (Column \texttt{ABSMAG\_I\_CORR}). It is derived from SDSS $i$-band \texttt{cmodel} magnitude and has been corrected for both foreground Galactic extinction and internal dust extinction due to inclination. For the internal correction, the authors used the $r$-band axial ratio ($b/a$) from SDSS exponential model fits (\texttt{expAB\_r}) and a simple logarithmic formula for the additional dust extinction due to inclination, $M_{i,\rm corr} = M_i + \gamma_i(M_i) \log(b/a)$. Because magnitudes uncorrected for inclination is desired for this study, I reversed the internal extinction correction using the listed $\gamma_i$ values in the catalog (\texttt{gamma\_i}) and the relation the authors used to calculate $\gamma_i$ from $M_i$: $\gamma_i = -0.15 M_i - 2.55$ for $M_i < -17$. For less luminous galaxies with $M_i > -17$, $M_i$ equals $M_{i,\rm corr}$ since $\gamma_i = 0$. After this process, the resulting $M_i$ magnitudes are corrected for foreground Galactic extinction only. 

Lastly, the uncertainties of the measurements are needed to construct the joint conditional PDF in Eq.\,\ref{eq:phi_r} \& \ref{eq:psi_r}. The mean uncertainty of \texttt{W50} in the catalog is 18\,\kms, which is comparable to the spectral resolution of ALFA (10\,\kms\ after Hanning smoothing, for a channel spacing of 5\,\kms). The uncertainty of \texttt{W20} is not reported because they are difficult to quantify for the adopted polynomial fitting algorithm, so I assume a conservative uncertainty of 20\,\kms\ for $W_{20}$. For the absolute magnitude $M_i$, I adopt the mean of the magnitude errors listed in \texttt{ABSMAG\_I\_CORR\_ERR}, which is 0.14\,mag.

Figure\,\ref{fig:phi} shows the distribution of the sample in the plane of $W_{20}$ and $M_i$. This 2D histogram represents $\tilde{\phi}(\tilde{x},\tilde{y})$ in Eq.\,\ref{eq:psi_r}.

\subsection{Conditional Probability Density Function} \label{sec:PDF}

\begin{figure}[!tb]
\epsscale{1.19}
\plotone{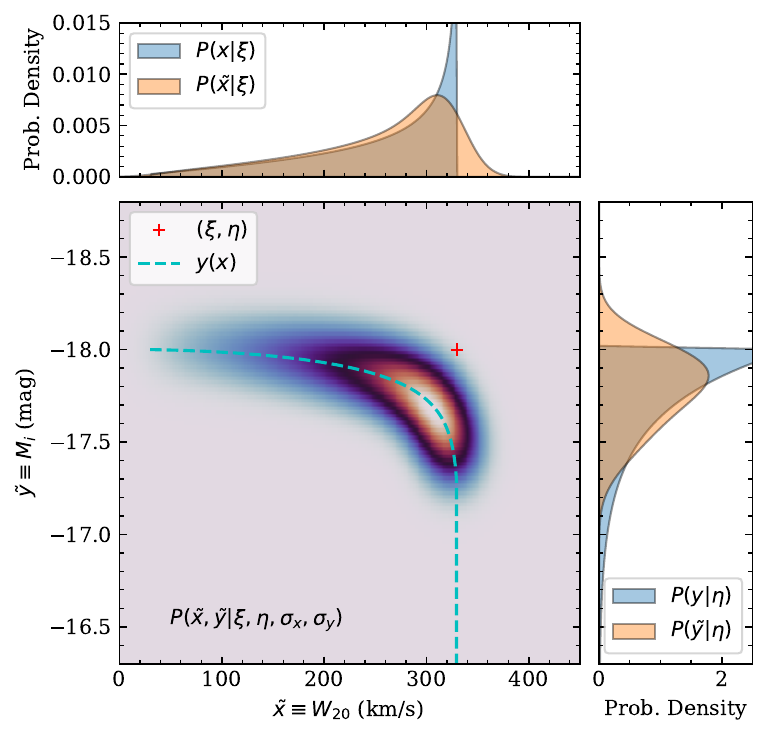}
\caption{The joint PDF $P(\tilde{x},\tilde{y} | \xi, \eta, \sigma_x, \sigma_y)$ for $\xi = 330$\,\kms, $\eta = -18$, $\sigma_x = 20$\,\kms, $\sigma_y = 0.14$, $\sigma_0 = 30$\,\kms, and $\gamma = 0.73$. The intrinsic values are indicated by the {\it red cross}, which is offset from the peak of the PDF. The top panel shows the PDF marginalized over the $\tilde{y}$-axis ({\it yellow}). This gives the PDF of the measured values, $P(\tilde{x}|\xi,\sigma_x)$ per Eq.\,\ref{eq:margin_y}, which can be compared with the PDF of the true projected values $P(x|\xi)$ from Eq.\,\ref{eq:fx} ({\it blue}) to see the effects of measurement errors. The right panel shows the joint PDF marginalized over the $\tilde{x}$-axis and $P(y|\eta)$ from Eq.\,\ref{eq:fy}. 
\label{fig:Pxy}} 
\epsscale{1.0}
\end{figure}

Both the \HI\ velocity width and the absolute magnitudes are affected by the inclination angle. The true projected values and the intrinsic values follow these simple relations:
\begin{align}
x &= \sqrt{(\xi \sin i)^2 + \sigma_0^2} \label{eq:x} \\
y &= \eta - \gamma \log(\cos i) \label{eq:y}
\end{align}
where the inclination angle $i$ is defined to be $0^\circ$ when viewed face-on and $90^\circ$ when viewed edge-on. 

In Eq.\,\ref{eq:x}, the projected velocity width is expressed as the quadrature sum of the line-of-sight projection of the edge-on velocity width and the velocity width from random motions ($\sigma_0$). I set $\sigma_0 = 30$\,\kms\ based on the lower boundary in the observed distribution of velocity widths shown in Figure\,\ref{fig:phi}. 

\begin{figure*}[!tb]
\epsscale{1.19}
\plotone{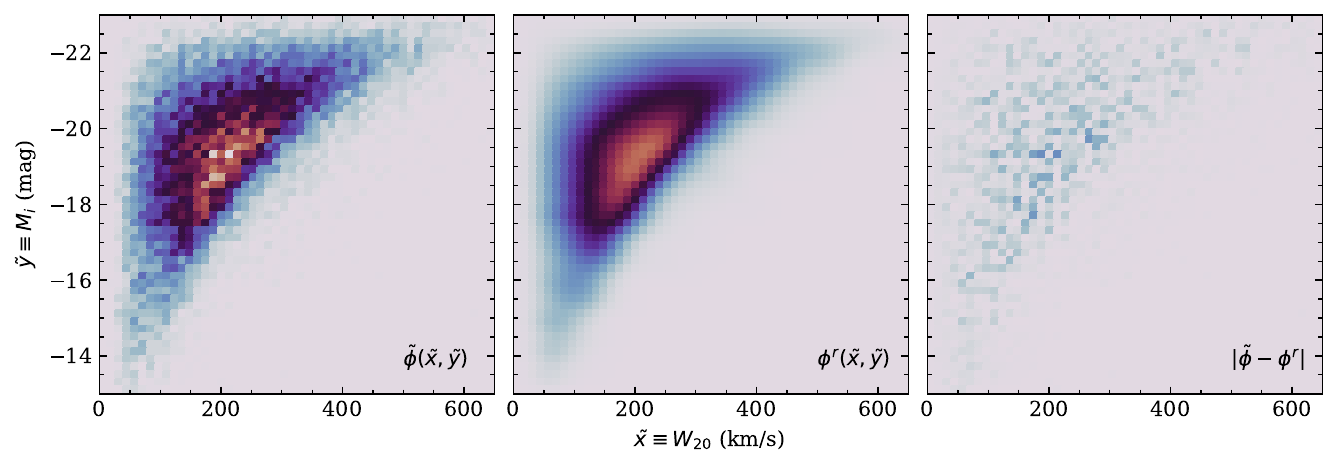}
\caption{Data vs. model. From left to right are respectively the observed $\tilde{\phi}$ distribution (same as Figure\,\ref{fig:phi}), the model $\phi^r$ distribution after 30 iterations, and the absolute differences between the two distributions. For fair comparison, the same color scale and contrast are used in all panels. 
\label{fig:fitqual}} 
\epsscale{1.0}
\end{figure*}

In Eq.\,\ref{eq:y}, the projected absolute magnitude follows the parameterization in Eq.\,27 of \citet{Giovanelli94}, which fits well the observed inclination dependency of $M^*$ (the knee of the optical luminosity function) for low-$z$ disk galaxies in the five SDSS filters \citep{Shao07}.  
Obviously, the extinction coefficient $\gamma$ depends on wavelength. Here I set $\gamma = 0.73$ using the result of \citet{Shao07} for the SDSS $i$-band ($\gamma_2$ in their Table 4). To understand the physical meaning of $\gamma$, one can compare the extinction term above, $A = \gamma \log(\sec i)$, with the plane-parallel extinction, $A^\prime = 2.5 \log(e) \tau_0 (\sec i-1)$, where $\tau_0$ is the face-on optical depth of the disk. At the face-on limit ($i \rightarrow 0$), $A \rightarrow \gamma \log(e) (\sec i-1)$, one sees that $\gamma = 2.5 \tau_0$. So $\gamma = 0.73$ implies an face-on optical depth of $\tau_0 = 0.29$, and disks become optically thick when $\gamma > 2.5$. 

Assuming the disks are randomly oriented on the sky, the PDF of the inclination angle is $P(i) = \sin i$. Given this and the relations in Eqs.\,\ref{eq:x}-\ref{eq:y}, the conditional PDF of $x$ and $y$ are:
\begin{align}
P(x|\xi) &= \frac{x/\xi}{\sqrt{\xi^2-(x^2-\sigma_0^2)}} {\rm \ when \ } 0 \leq x \leq \xi \label{eq:fx} \\
P(y|\eta) &= \frac{\ln 10}{\gamma} 10^{(\eta-y)/\gamma} {\rm \ when \ } \eta < y \label{eq:fy}
\end{align}
In addition, since both $x$ and $y$ are related to the same inclination angle $i$, the two are correlated: 
\begin{equation}
y = \eta - 0.5 \gamma \log(1-\frac{x^2-\sigma_0^2}{\xi^2}) \label{eq:yx}
\end{equation}

The joint conditional PDF of $x$ and $y$ is determined by Eqs.\,\ref{eq:fx}-\ref{eq:yx} because  when integrated over one axis it must recover the conditional PDF of the other axis:
\begin{align}
\int P(x,y|\xi,\eta) dy &= P(x|\xi) = \frac{x/\xi}{\sqrt{\xi^2-(x^2-\sigma_0^2)}} \label{eq:margin_y} \\
\int P(x,y|\xi,\eta) dx &= P(y|\eta) = \frac{\ln 10}{\gamma} 10^{(\eta-y)/\gamma} \label{eq:margin_x}
\end{align}
The correlation relation in Eq.\,\ref{eq:yx} explains why $\eta$ drops off after the integral over $y$ in Eq.\,\ref{eq:margin_y} and $\xi$ drops off in Eq.\,\ref{eq:margin_x}: Eq.\,\ref{eq:yx} allows $\eta$ or $\xi$ to be expressed by the other three parameters. 

All measurements have errors, and the errors scatter the measured value around the true value following a PDF that is usually assumed to be Gaussian. Because the velocity widths and the absolute magnitudes come from two different surveys (ALFALFA and SDSS), and only the latter depends on the distance to the source, we can safely assume that the measurement errors in $x$ and $y$ are uncorrelated. As a result, the error PDF is a 2D Gaussian with major and minor axes aligned with the $x$ and $y$ axes:
\begin{equation}
G(x-\tilde{x},y-\tilde{y}|\sigma_x,\sigma_y) = \frac{1}{2\pi \sigma_x \sigma_y} \exp \Big(-\frac{(\tilde{x}-x)^2}{2\sigma_x^2} \Big) \exp \Big(-\frac{(\tilde{y}-y)^2}{2\sigma_y^2} \Big)
\end{equation}
And the joint conditional PDF of $\tilde{x}$ and $\tilde{y}$ is then $P(x,y|\xi,\eta)$ convolved with the 2D Gaussian: 
\begin{equation}
P(\tilde{x},\tilde{y} | \xi, \eta, \sigma_x, \sigma_y) = \iint P(x,y|\xi,\eta) G(x-\tilde{x},y-\tilde{y}|\sigma_x,\sigma_y) dx dy
\end{equation}
Efficient convolution algorithms based on Fast Fourier Transform (FFT) can be used to evaluate $P(\tilde{x},\tilde{y} | \xi, \eta, \sigma_x, \sigma_y)$ on the $(\tilde{x},\tilde{y})$ plane for a grid of $(\xi,\eta)$. Over the parameter ranges covered by the data, $0 < x < 650$\,\kms\ and $-23 < y < -13$, I calculate the joint PDF on a 51$\times$51 grid in both ($\tilde{x},\tilde{y}$) and ($\xi, \eta$) with spacings of 13\,\kms\ and 0.2\,mag. The resulting 4D array can be interpolated and used for integrations in the iterative process (Eqs.\,\ref{eq:phi_r} and \ref{eq:psi_r}). 

Figure\,\ref{fig:Pxy} shows an example of the joint PDF. As expected, the population is dominated by more inclined disks, which show higher velocity widths but suffer more internal dust extinction. The correlation between $\tilde{x}$ and $\tilde{y}$ in Eq.\,\ref{eq:yx} is also evident in the plot. As described in \S\ref{sec:phi}, I estimated the mean measurement errors to be $\sigma_x = 20$\,\kms\ and $\sigma_y = 0.14$\,mag. Note that an accurate knowledge of the measurement errors is important to quantify the intrinsic dispersion of the restored intrinsic relation, because if the errors were underestimated (overestimated), the recovered intrinsic distribution would have shown a larger (smaller) scatter. 

\subsection{Rectified Intrinsic Distribution} \label{sec:psi}

To start the iterative process, one needs to prescribe an initial distribution for $\psi(\xi,\eta)$. Usually it is recommended to prescribe the observed distribution as the initial guess, $\psi^0(\xi,\eta) = \tilde{\phi}(\tilde{x},\tilde{y})$, to speed up the convergence. But because the position of the intrinsic position is offset from the peak of the joint PDF, as illustrated in Figure\,\ref{fig:Pxy}, one would expect similar offsets between $\psi$ and $\phi$. So I simply prescribed a flat distribution as the initial guess, $\psi^0(\xi,\eta) = {\rm constant}$, over the parameter ranges covered by the data. 

At the end of each iteration, both the rectified distribution $\psi^r(\xi,\eta)$ (Eq.\,\ref{eq:psi_r}) and its corresponding projected distribution $\phi^r(\tilde{x},\tilde{y})$ (Eq.\,\ref{eq:phi_r}) are produced. The latter can be directly compared with the observed distribution $\tilde{\phi}(\tilde{x},\tilde{y})$ to assess the improvement of the model after each iteration. As already noted in previous works, the Lucy algorithm is very efficient. After just a few iterations, a narrow curved distribution begins to emerge in $\psi^r$ and the resulting $\phi^r$ starts to converge onto the input $\tilde{\phi}$. Based on the residual map ($\tilde{\phi} - \phi^r$) and the expected Poisson noise of $\tilde{\phi}$, I find that the reduced $\chi^2_\nu$ decreases from 13.7 after the first iteration to 1.0 after 30 iterations, which is a natural point to stop. Figure\,\ref{fig:fitqual} shows that the model distribution of the observables accurately reproduces the observed distribution, and there is only statistical noise left in the residual map. The following results are all based on the products after 30 iterations.

Figure\,\ref{fig:psi_corner} shows the rectified distribution. The main panel shows that $\psi^r(\xi,\eta)$ is confined to a narrow, continuous sequence along the diagonal direction, revealing a tight correlation between the {\it edge-on} \HI\ line width and the {\it face-on} $i$-band absolute magnitude of \HI-selected galaxies. This ``ridge'' is the $i$-band TFR from the full sample of 28,264 \HI-selected nearby ($z < 0.06$) galaxies in the ALFALFA-SDSS catalog. Note that because the full sample is used (as opposed to selecting only high inclination disks as in previous studies), the absence of any significant secondary trend in the intrinsic distribution shows that the overwhelming majority of \HI-selected galaxies follow a single correlation. 

The TFR is usually parameterized as a power law between luminosity and velocity-width: $L \propto W^\beta$. Since magnitudes are used here, the relation translates to:
\begin{equation}
M_{i,\rm face-on} = M_0 - 2.5 \beta \log\Big(\frac{W_{\rm 20,edge-on}}{250\,\,{\rm km/s}} \Big)
\label{eq:TFR}
\end{equation}
where $\beta$ is the power-law slope and $M_0$ is the absolute magnitude at 250\,\kms. To determine these parameters, I first measure the width of the ridge and the location of its peak at each absolute magnitude by fitting a Gaussian function along the velocity width direction, and then I find the best-fit power-law model by minimizing the residuals along the velocity direction, using the Gaussian $\sigma$ widths as relative errors. Because the fit utilizes errors along the axis of velocity width (as opposed to the magnitude axis or both axes), the result is often called the ``inverse'' TFR \citep{Tully00}. The benefit of this approach is that it minimizes the Malmquist bias in luminosity, so it is preferred for studies of galaxy distances (a key application of the TFR). 

The best-fit power law to the ridge in the restored distribution has $\beta = 4.39\pm0.06$ and $M_0 = -19.77\pm0.04$, which is shown as a red dashed curve in Figure\,\ref{fig:psi_corner}. Nearly the same TFR ($\beta = 4.26\pm0.07$ and $M_0 = -19.68\pm0.04$) is obtained using the conventional method, where individual galaxy's velocity width and absolute magnitude are corrected using the inclination angle estimated from its axial ratio (see \S\,\ref{sec:comparison}). It should be noted that the absolute calibration of the TFR requires zero-point-calibrator galaxies that have distances from standard candles like the Cepheid period-luminosity relation and the tip of the red giant branch. Because the distances to the ALFALFA galaxies were derived from the Hubble-Lema\^{i}tre law with an assumed Hubble constant $H_0 = 70$\,\kms\,Mpc$^{-1}$, the goal here is not to provide an absolute calibration of the TFR, but to demonstrate the capabilities of the statistical rectification method. 

For comparison, the Cosmicflows-4 TFR in $i$-band has $\beta = 3.33\pm0.05$ and $M_0 = -20.01\pm0.10$ \citep{Kourkchi20a}\footnote{Converted from parameters $Slope$, $ZP$, and $C_{\rm zp}$ listed in their Table 2.}. The discrepancies between the ALFALFA TFR and the Cosmicflows-4 TFR are likely due to the differences in the techniques applied to measure and adjust velocity widths, apparent magnitudes, and distances. A proper investigation of this issue requires using the Cosmicflows-4 calibration sample that includes 648 slope-calibrator cluster galaxies and 94 zero-point-calibrator galaxies. Because the small sample would lead to a noisy input distribution function $\tilde{\phi}$, it might be preferable to adopt a parameterized maximum likelihood approach similar to that of \citet{Isbell18}\footnote{\url{https://github.com/fuhaiastro/IXF18} provides a Python notebook of the method.}, rather than the non-parameterized rectification method described here. In the parameterized approach, the inclination angle will still be treated as a random variable with a known PDF instead of a parameter that needs to be measured from the axial ratio. The parameters of the TFR and their uncertainties will be constrained by maximizing the likelihood of the data given the model with a Markov chain Monte Carlo (MCMC) sampler. In addition, the PDF of the inclination angle needs to be modified to account for the exclusion of disk galaxies with inclination angles less than 45$^\circ$ in the Cosmicflows-4 sample.

A closer inspection of the ridge in the rectified distribution reveals significant deviations from a single power law at the brighter end. As shown in the inset of Figure\,\ref{fig:psi_corner}, the departure from the best-fit power law becomes evident above $-22$\,mag, where the slope flattens to $\beta \sim 3.3$. Similar curvatures in TFRs have been reported previously: e.g., the slope of the Cosmicflows-4 TFR in $i$-band also decreases above $-22$\,mag, where the TFR is better fit by a quadratic function of $\log W$ than a linear function \citep{Kourkchi20a}. The curvature could be at least partially explained by the \HI-selection bias. Because gas-rich sub-luminous galaxies are preferentially selected than gas-poor luminous galaxies, this bias steepens the luminosity$-$line width correlation towards the fainter end. By adding the ``dark'' gas mass to the ``luminous'' stellar mass, the curvature of the correlation could be minimized, as demonstrated in previous studies of the baryonic TFR \citep[e.g.,][]{McGaugh00,McGaugh05,Lelli19,Kourkchi22}.

The width of the restored distribution is a measure of the intrinsic scatter of the TFR. In the velocity direction, the logarithmic Gaussian width is roughly constant at $M_i < -17$, with a median of $\sigma(\log W_{20}) = 0.06$\,dex. On the other hand, the Gaussian width in the magnitude direction $\sigma(M_i)$ decreases from $\sim$0.8\,mag to $\sim$0.4\,mag as $W_{20}$ increases from $\sim$150\,\kms\ to $\sim$450\,\kms; This is similar to the TFR scatter found in \citet{Kourkchi20a}. Note that these Gaussian widths of the ridge should be considered as the upper limits on the intrinsic dispersion of the TFR, because there could be additional measurement errors that are not included in the joint PDF built in \S\,\ref{sec:PDF}.

Finally, there are two important by-products from this exercise. Marginalized over each axis, the rectified distribution $\psi^r(\xi,\eta)$ provides the edge-on \HI\ velocity-width function and the face-on $i$-band luminosity function of \HI-selected disk galaxies. Of course, both functions remain uncorrected for the detection incompleteness and the luminosity-dependent volume incompleteness. Nevertheless, the effects of inclination correction is evident when comparing the histograms in the side panels of Figure\,\ref{fig:psi_corner}. The face-on luminosity function is simply shifted by $\sim$0.2\,mag towards to brighter end. The changes in the velocity-width function are more pronounced. After the rectification, not only its peak shifts by $\sim$50\,\kms\ to the higher end, but also its slopes on both sides of the peak become steeper. 

\subsection{Comparison with conventional method} \label{sec:comparison}

\begin{figure}[!tb]
\epsscale{1.19}
\plotone{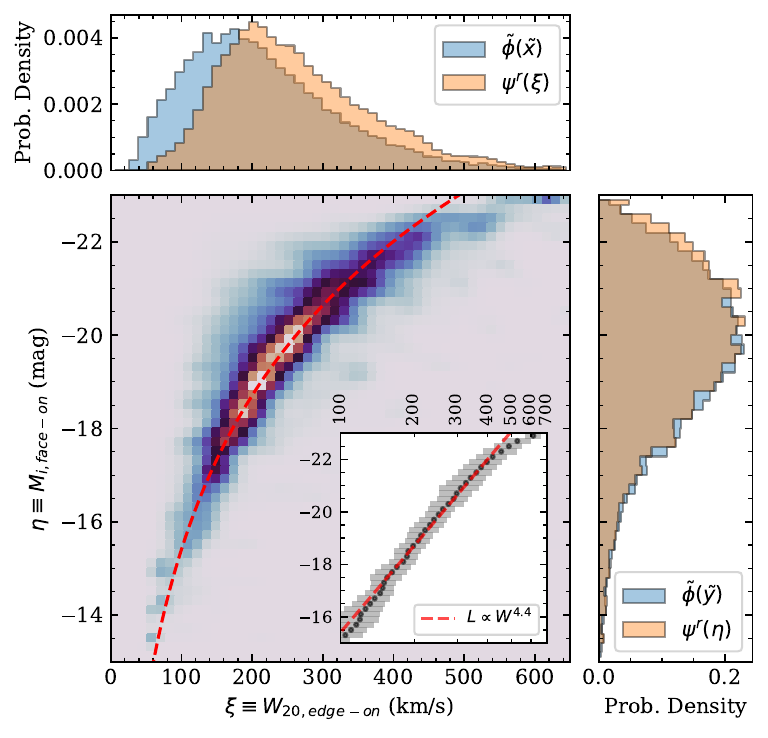}
\caption{The rectified distribution $\psi^r(\xi,\eta)$ after 30 iterations. This is the restored $i$-band TFR. The red dashed curve in the main panel shows the best-fit power law. The inset shows the curvature of the relation by comparing the peak positions of the ridge ({\it black circles}) with the same power law. The gray shaded areas show the Gaussian $\sigma$ widths of the ridge {\it in velocity}. The side panels compare the marginalized distributions before and after rectification ({\it blue} and {\it yellow}, respectively). They illustrates the differences between the observed \HI-velocity width function and $i$-band luminosity function and the rectified functions. Note that selection effects such as the magnitude limit and the volume incompleteness have not been corrected.
\label{fig:psi_corner}} 
\epsscale{1.0}
\end{figure} 

\begin{figure}[!tb]
\epsscale{1.19}
\plotone{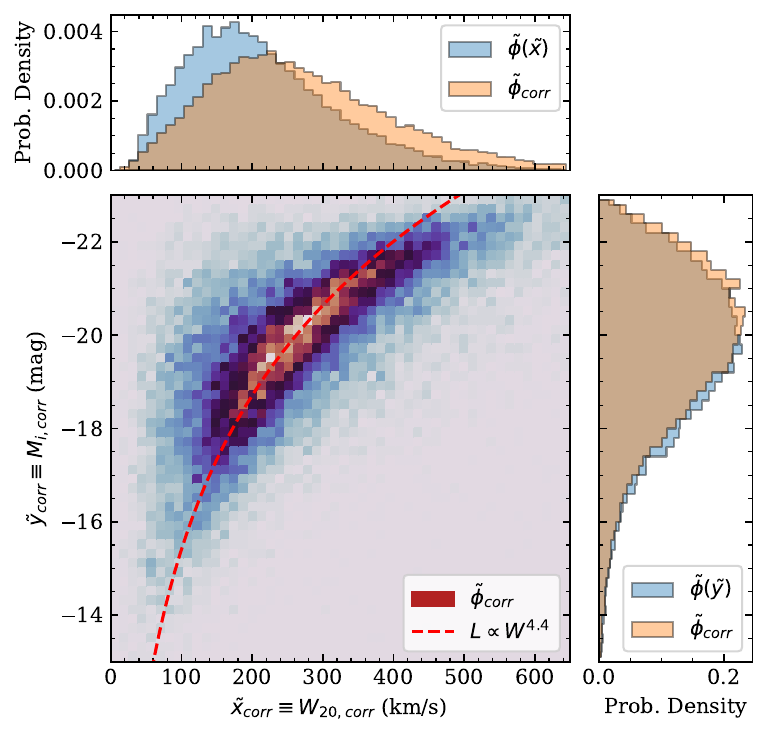}
\caption{The distribution of the ALFALFA-SDSS galaxy sample after individual inclination angle correction based on the axial ratio ($b/a$). The red dashed curve shows the best-fit power law in Figure\,\ref{fig:psi_corner}. The side panels compare the marginalized distributions before and after the inclination-angle correction. 
\label{fig:phi_corr}} 
\epsscale{1.0}
\end{figure} 

Having successfully restored the TFR through rectification, in this subsection I compare the results with those from the traditional method of $b/a$-based individual inclination correction. The procedure is straight-forward. First, one estimates $\sin i$ of each galaxy using the following equation from \citet{Hubble26}:
\begin{equation}
\sin^2 i = \frac{1-(b/a)^2}{1-q_0^2} \label{eq:hubble26}
\end{equation}
where $q_0$ is the {\it assumed} edge-on axial ratio of the disk. Next, one corrects the projection effects in both the observed velocity width and the observed absolute magnitude using the estimated inclination angle of each galaxy and the following relations (same as Eqs.\,\ref{eq:x} and \ref{eq:y}):
\begin{align}
W_{\rm 20,corr} &= \sqrt{W_{20}^2-\sigma_0^2}/\sin i \nonumber \\
M_{i,\rm corr} &= M_i + \gamma \log(\cos i) \label{eq:corr}
\end{align} 
Finally, one generates the 2D histogram using the inclination-corrected measurements. 

For a fair comparison, I carry out the $b/a$-based inclination corrections to the same ALFALFA-SDSS sample. To estimate the inclination angles, I assume $q_0 = 0.15$ and adopt the $r$-band axial ratio from SDSS exponential model fits (Column \texttt{expAB\_r}). When applying the corrections, I adopt $\sigma_0 = 30$\,\kms\ and $\gamma = 0.73$ to be consistent with the rectification analysis. Figure\,\ref{fig:phi_corr} shows the resulting distribution using the corrected velocity widths and absolute magnitudes. This figure should be directly compared with Figure\,\ref{fig:psi_corner} to see the significant differences between the results from the two different methods. There are substantial populations of ``outlier" galaxies that seem to be either under-corrected or over-corrected in the traditional method. Due to these outliers, the corrected TFR appears much broader than that from the rectification method, with median Gaussian widths of $\sigma(\log W_{20}) = 0.13$\,dex and $\sigma(M_i) = 0.74$\,mag. Nevertheless, similar slope and intercept are obtained by fitting the ridge with a power law ($\beta = 4.26\pm0.07$ and $M_0 = -19.68\pm0.04$) and a similar curvature is seen at the bright end.

The comparison between Figure\,\ref{fig:psi_corner} and Figure\,\ref{fig:phi_corr} shows that the statistical rectification method produces better results than the traditional method of individual inclination correction based on axial ratio, when both methods are applied to the full sample of ALFALFA galaxies. Now I briefly discuss why this is the case. When compared with the statistical rectification method, the main disadvantages of the traditional method include:
\begin{enumerate}
\item One additional measurement set, e.g., the axial ratio $b/a$, must be used to estimate $\sin i$;
\item The estimated inclination angle are not always reliable, introducing additional errors to the data;
\item Measurement errors of the observables are not corrected for.
\end{enumerate}
The first item is particularly problematic at high redshifts when the angular sizes of galaxies are small compared to the spatial resolution. The latter two items make the resulting TFR broader, which in turn makes it more difficult to quantify the intrinsic scatter of the relation. 

Why the estimated inclination angles are unreliable? Eq.\,\ref{eq:hubble26} was first derived by \citet{Hubble26}, who assumed disk galaxies were axisymmetric oblate ellipsoids. One faces three main problems when using this equation to estimate the inclination angle: 
\begin{enumerate}
\item The edge-on axial ratio, $q_0$, is not well determined and likely varies with morphological type and luminosity; the literature has assumed a range of values between $0.10 \leq q_0 \leq 0.25$ \citep[e.g.,][]{Giovanelli94, Xilouris99, Ubler17}.
\item Disk galaxies are not axisymmetric. Instead, they show median ellipticity between $0.07 \leq \epsilon \leq 0.18$ \citep[e.g.,][]{Ryden06}.
\item Axial ratios from different methods differ (e.g., morphological fitting vs. isophotes) and are affected by observational conditions.
\end{enumerate}
For these reasons, previous TFR studies have excluded galaxies with low inclination angles to minimize the amount of correction to the velocity widths (e.g., $\sin i > 0.87$ when $i > 60^\circ$). But this exclusion alone would severely reduce the sample size; e.g., the ALFALFA-SDSS galaxy sample would be reduced by a factor of three when I exclude galaxies with $i < 60^\circ$. This exclusion not only reduces the statistical accuracy of the result, but also made it impossible to assess whether the excluded sample follows the same intrinsic correlation as the included sample. 

\section{Summary and Future Prospect} \label{sec:summary}

The TFR is an important empirical correlation between the edge-on rotation velocity and the face-on luminosity of disk galaxies. To determine this relation, three sets of observed properties are typically required: galaxy-integrated line widths ($W$), absolute magnitudes ($M$), and the axial ratios ($b/a$). The axial ratios are needed because the first two observed properties needed to be corrected for the inclination angle ($i$) of the disk. In this work, I have demonstrated a rectification method that replaces {\it individual} inclination correction with {\it ensemble} statistical correction. It determines the TFR with only two sets of observables ($W$ and $M$) and utilizes the full sample of disk galaxies regardless of their inclination angles. The general philosophy of the method is as follows. When the observed properties can be converted from the intrinsic properties and the inclination angles with some known relations, one can predict joint PDF of the observed properties by assuming randomly oriented disks and random measurement errors. The recovery of the distribution of the intrinsic properties from the distribution of observables then becomes a reversal of {\it the law of total probability} in Eq.\,\ref{eq:general} that can be tackled numerically with the iterative rectification algorithm of \citet{Lucy74}.

The statistical rectification method will be particularly useful for TFR studies at high redshift, where it is difficult to estimate the inclination angles because of limited spatial resolution. Here I have demonstrated the method with a low-redshift galaxy sample because \HI\ measurements are available and the data set allows a direct comparison with the conventional method of individual inclination correction. With 28,264 \HI-detected disk galaxies from the ALFALFA-SDSS survey at $z < 0.06$, I show that the rectified distribution of {\it edge-on} \HI\ line width and {\it face-on} $i$-band absolute magnitude of \HI-selected disk galaxies follows a sharp power-law relation, $M_i = M_0 - 2.5 \beta \ [\log(W_{\rm 20, HI}/250 {\rm km/s})]$, with $\beta = 4.39\pm0.06$ and $M_0 = -19.77\pm0.04$. The intrinsic dispersion of the TFR in velocity width is almost constant, with $\sigma(\log W_{20}) \lesssim 0.06$\,dex between $-23 < M_i < -17$, while the dispersion in absolute magnitude, $\sigma({M_i})$, decreases from $\sim$0.8\,mag among slow rotators to $\sim$0.4\,mag among fast rotators. 
A closer examination of the restored TFR reveals significant deviations from a single power law at the brighter end, confirming previous studies. 
The absence of any significant secondary trends in the rectified distribution shows that essentially all \HI-selected disk galaxies follow a single TFR. In addition, the rectified distribution marginalized over each axis provides the inclination-corrected \HI\ velocity width function and the luminosity function of these galaxies, both of which show significant changes from the uncorrected distributions.

Moving forward, the method can be applied to investigate the baryonic TFR of ALFALFA galaxies, by converting the available \HI\ flux and optical photometry to gas mass and stellar mass. To make the results comparable to the Cosmicflows-4 baryonic TFR \citep{Kourkchi22}, special attention must be paid to \HI\ velocity width measurements, magnitude measurements and corrections, mass-to-light ratios, molecular-to-atomic gas ratios, and degeneracies among the suite of modeled parameters. A number of important biases should also be quantified and corrected with the help of synthetic data sets: e.g., the residual Malmquist bias from the increasing luminosity limit at greater distances, asymmetric magnitude scattering when the number density of galaxies decreases exponentially with luminosity, the bright-end curvature of the TFR, and the \HI\ flux limit that preferentially select more gas-rich galaxies at greater distances \citep{Kourkchi20b,Kourkchi22}. Of course, the same biases affect the $i$-band TFR in this work. The resulting baryonic TFR can be calibrated onto the absolute scale using spiral galaxies with standard candle distances from the Cepheid period-luminosity relation and the tip of the red giant branch, and the required vertical offset would provide a constraint on the Hubble constant $H_0$ (although the result is fundamentally limited by the systematic uncertainties of the distance calibrators).

The method has a wide range of applicability in observational astronomy. It can be applied to any observational problems where (1) the observed properties can be converted from the intrinsic properties by a relation that involves a ``hidden'' parameter (e.g., $i$) and (2) the PDF of the hidden parameter is precisely known. For example, in future studies of disk galaxies beyond the TFR, the method can be used to restore the distributions of edge-on disk thickness ($q_0$) and face-on ellipticity ($\epsilon$)  as functions of face-on absolute magnitude (i.e., extending the work of \citealt{Binney81, Vincent05, Ryden06, Roychowdhury10}). The potential of the method grows stronger with the increasing number of astronomical surveys generating statistical datasets encompassing large samples of galaxies near and far. 

\acknowledgements

I thank the referee, Brent Tully, for providing comments that improved the {\it Letter}. I also appreciate the discussions with my colleagues Steve Spangler, Ken Gayley, and Kevin Hall. This work is supported by the National Science Foundation (NSF) grant AST-2103251.

\bibliographystyle{apj}
\bibliography{draft}

\begin{thebibliography}{32}
\expandafter\ifx\csname natexlab\endcsname\relax\def\natexlab#1{#1}\fi

\bibitem[{{Ball} {et~al.}(2023){Ball}, {Haynes}, {Jones}, {Peng}, {Durbala},
  {Koopmann}, {Ribaudo}, \& {O'Donoghue}}]{Ball23}
{Ball}, C.~J., {Haynes}, M.~P., {Jones}, M.~G., {et~al.} 2023, \apj, 950, 87

\bibitem[{{Binney} \& {de Vaucouleurs}(1981)}]{Binney81}
{Binney}, J., \& {de Vaucouleurs}, G. 1981, \mnras, 194, 679

\bibitem[{{Cornwell}(2009)}]{Cornwell09}
{Cornwell}, T.~J. 2009, \aap, 500, 65

\bibitem[{{Desmond}(2017)}]{Desmond17}
{Desmond}, H. 2017, \mnras, 472, L35

\bibitem[{{Durbala} {et~al.}(2020){Durbala}, {Finn}, {Crone Odekon}, {Haynes},
  {Koopmann}, \& {O'Donoghue}}]{Durbala20}
{Durbala}, A., {Finn}, R.~A., {Crone Odekon}, M., {et~al.} 2020, \aj, 160, 271

\bibitem[{{Giovanelli} {et~al.}(1994){Giovanelli}, {Haynes}, {Salzer},
  {Wegner}, {da Costa}, \& {Freudling}}]{Giovanelli94}
{Giovanelli}, R., {Haynes}, M.~P., {Salzer}, J.~J., {et~al.} 1994, \aj, 107,
  2036

\bibitem[{{Haynes} {et~al.}(2018){Haynes}, {Giovanelli}, {Kent}, {Adams},
  {Balonek}, {Craig}, {Fertig}, {Finn}, {Giovanardi}, {Hallenbeck}, {Hess},
  {Hoffman}, {Huang}, {Jones}, {Koopmann}, {Kornreich}, {Leisman}, {Miller},
  {Moorman}, {O'Connor}, {O'Donoghue}, {Papastergis}, {Troischt}, {Stark}, \&
  {Xiao}}]{Haynes18}
{Haynes}, M.~P., {Giovanelli}, R., {Kent}, B.~R., {et~al.} 2018, \apj, 861, 49

\bibitem[{{H{\"o}gbom}(1974)}]{Hogbom74}
{H{\"o}gbom}, J.~A. 1974, \aaps, 15, 417

\bibitem[{{Hubble}(1926)}]{Hubble26}
{Hubble}, E.~P. 1926, \apj, 64, 321

\bibitem[{{Isbell} {et~al.}(2018){Isbell}, {Xue}, \& {Fu}}]{Isbell18}
{Isbell}, J.~W., {Xue}, R., \& {Fu}, H. 2018, \apjl, 869, L37

\bibitem[{{Kourkchi} {et~al.}(2020{\natexlab{a}}){Kourkchi}, {Tully}, {Anand},
  {Courtois}, {Dupuy}, {Neill}, {Rizzi}, \& {Seibert}}]{Kourkchi20a}
{Kourkchi}, E., {Tully}, R.~B., {Anand}, G.~S., {et~al.} 2020{\natexlab{a}},
  \apj, 896, 3

\bibitem[{{Kourkchi} {et~al.}(2022){Kourkchi}, {Tully}, {Courtois}, {Dupuy}, \&
  {Guinet}}]{Kourkchi22}
{Kourkchi}, E., {Tully}, R.~B., {Courtois}, H.~M., {Dupuy}, A., \& {Guinet}, D.
  2022, \mnras, 511, 6160

\bibitem[{{Kourkchi} {et~al.}(2020{\natexlab{b}}){Kourkchi}, {Tully},
  {Eftekharzadeh}, {Llop}, {Courtois}, {Guinet}, {Dupuy}, {Neill}, {Seibert},
  {Andrews}, {Chuang}, {Danesh}, {Gonzalez}, {Holthaus}, {Mokelke}, {Schoen},
  \& {Urasaki}}]{Kourkchi20b}
{Kourkchi}, E., {Tully}, R.~B., {Eftekharzadeh}, S., {et~al.}
  2020{\natexlab{b}}, \apj, 902, 145

\bibitem[{{Lelli} {et~al.}(2019){Lelli}, {McGaugh}, {Schombert}, {Desmond}, \&
  {Katz}}]{Lelli19}
{Lelli}, F., {McGaugh}, S.~S., {Schombert}, J.~M., {Desmond}, H., \& {Katz}, H.
  2019, \mnras, 484, 3267

\bibitem[{{Lucy}(1974)}]{Lucy74}
{Lucy}, L.~B. 1974, \aj, 79, 745

\bibitem[{{McGaugh}(2005)}]{McGaugh05}
{McGaugh}, S.~S. 2005, \apj, 632, 859

\bibitem[{{McGaugh} {et~al.}(2000){McGaugh}, {Schombert}, {Bothun}, \& {de
  Blok}}]{McGaugh00}
{McGaugh}, S.~S., {Schombert}, J.~M., {Bothun}, G.~D., \& {de Blok}, W.~J.~G.
  2000, \apjl, 533, L99

\bibitem[{{Orieux} {et~al.}(2010){Orieux}, {Giovannelli}, \&
  {Rodet}}]{Orieux10}
{Orieux}, F., {Giovannelli}, J.-F., \& {Rodet}, T. 2010, Journal of the Optical
  Society of America A, 27, 1593

\bibitem[{{Richardson}(1972)}]{Richardson72}
{Richardson}, W.~H. 1972, Journal of the Optical Society of America
  (1917-1983), 62, 55

\bibitem[{{Roychowdhury} {et~al.}(2010){Roychowdhury}, {Chengalur}, {Begum}, \&
  {Karachentsev}}]{Roychowdhury10}
{Roychowdhury}, S., {Chengalur}, J.~N., {Begum}, A., \& {Karachentsev}, I.~D.
  2010, \mnras, 404, L60

\bibitem[{{Ryden}(2006)}]{Ryden06}
{Ryden}, B.~S. 2006, \apj, 641, 773

\bibitem[{{Shao} {et~al.}(2007){Shao}, {Xiao}, {Shen}, {Mo}, {Xia}, \&
  {Deng}}]{Shao07}
{Shao}, Z., {Xiao}, Q., {Shen}, S., {et~al.} 2007, \apj, 659, 1159

\bibitem[{{Springob} {et~al.}(2005){Springob}, {Haynes}, {Giovanelli}, \&
  {Kent}}]{Springob05}
{Springob}, C.~M., {Haynes}, M.~P., {Giovanelli}, R., \& {Kent}, B.~R. 2005,
  \apjs, 160, 149

\bibitem[{{Tiley} {et~al.}(2016){Tiley}, {Bureau}, {Saintonge}, {Topal},
  {Davis}, \& {Torii}}]{Tiley16}
{Tiley}, A.~L., {Bureau}, M., {Saintonge}, A., {et~al.} 2016, \mnras, 461, 3494

\bibitem[{{Topal} {et~al.}(2018){Topal}, {Bureau}, {Tiley}, {Davis}, \&
  {Torii}}]{Topal18}
{Topal}, S., {Bureau}, M., {Tiley}, A.~L., {Davis}, T.~A., \& {Torii}, K. 2018,
  \mnras, 479, 3319

\bibitem[{{Tully} \& {Courtois}(2012)}]{Tully12}
{Tully}, R.~B., \& {Courtois}, H.~M. 2012, \apj, 749, 78

\bibitem[{{Tully} \& {Fisher}(1977)}]{Tully77}
{Tully}, R.~B., \& {Fisher}, J.~R. 1977, \aap, 54, 661

\bibitem[{{Tully} \& {Pierce}(2000)}]{Tully00}
{Tully}, R.~B., \& {Pierce}, M.~J. 2000, \apj, 533, 744

\bibitem[{{{\"U}bler} {et~al.}(2017){{\"U}bler}, {F{\"o}rster Schreiber},
  {Genzel}, {Wisnioski}, {Wuyts}, {Lang}, {Naab}, {Burkert}, {van Dokkum},
  {Tacconi}, {Wilman}, {Fossati}, {Mendel}, {Beifiori}, {Belli}, {Bender},
  {Brammer}, {Chan}, {Davies}, {Fabricius}, {Galametz}, {Lutz}, {Momcheva},
  {Nelson}, {Saglia}, {Seitz}, \& {Tadaki}}]{Ubler17}
{{\"U}bler}, H., {F{\"o}rster Schreiber}, N.~M., {Genzel}, R., {et~al.} 2017,
  \apj, 842, 121

\bibitem[{{Vincent} \& {Ryden}(2005)}]{Vincent05}
{Vincent}, R.~A., \& {Ryden}, B.~S. 2005, \apj, 623, 137

\bibitem[{{Xilouris} {et~al.}(1999){Xilouris}, {Byun}, {Kylafis}, {Paleologou},
  \& {Papamastorakis}}]{Xilouris99}
{Xilouris}, E.~M., {Byun}, Y.~I., {Kylafis}, N.~D., {Paleologou}, E.~V., \&
  {Papamastorakis}, J. 1999, \aap, 344, 868

\bibitem[{{Zaritsky} {et~al.}(2014){Zaritsky}, {Courtois}, {Mu{\~n}oz-Mateos},
  {Sorce}, {Erroz-Ferrer}, {Comer{\'o}n}, {Gadotti}, {Gil de Paz}, {Hinz},
  {Laurikainen}, {Kim}, {Laine}, {Men{\'e}ndez-Delmestre}, {Mizusawa}, {Regan},
  {Salo}, {Seibert}, {Sheth}, {Athanassoula}, {Bosma}, {Cisternas}, {Ho}, \&
  {Holwerda}}]{Zaritsky14a}
{Zaritsky}, D., {Courtois}, H., {Mu{\~n}oz-Mateos}, J.-C., {et~al.} 2014, \aj,
  147, 134

\end{thebibliography}

\end{document}